\documentclass[3p]{elsarticle}
\makeatletter
\def\@author#1{\g@addto@macro\elsauthors{\normalsize%
    \def\baselinestretch{1}%
    \upshape\authorsep#1\unskip\textsuperscript{%
      \ifx\@fnmark\@empty\else\unskip\sep\@fnmark\let\sep=,\fi
      \ifx\@corref\@empty\else\unskip\sep\@corref\let\sep=,\fi
      }%
    \def\authorsep{\unskip,\space}%
    \global\let\@fnmark\@empty
    \global\let\@corref\@empty  
    \global\let\sep\@empty}%
    \@eadauthor={#1}
}
\makeatother
\usepackage{graphicx, bm, balu, caption, subcaption, booktabs}
\usepackage{amsmath, amsfonts}
\usepackage{algorithm2e}

\DeclareMathOperator*{\argmin}{argmin} 
\newcommand{\ufo}{\ensuremath{\vv u^{W\mkern-5mu R}}}
\newcommand{\ffo}{\ensuremath{\vv f^{W\mkern-5mu R}}}
\newcommand{\qfo}{\ensuremath{\vv q^{W\mkern-5mu R}}}
\newcommand{\uro}{\ensuremath{\vv u^{R\mkern-1mu O}}}
\newcommand{\fro}{\ensuremath{\vv f^{R\mkern-1mu O}}}
\newcommand{\qro}{\ensuremath{\vv q^{R\mkern-1mu O}}}
\newcommand{\TRANS}{\ensuremath{\vv q^{R\mkern-1mu O}}}
\newcommand{\TDNS}{\ensuremath{\vv q^{W\mkern-5mu R}}}
\newcommand{\FDNS}{\ensuremath{\vv F^{W\mkern-5mu R}}}
\newcommand{\qdns}{\ensuremath{\bm{q}_i^{D\!N\!S}}}
\newcommand{\qrans}{\ensuremath{\bm{q}_i^{R\!A\!N\!S}}}
\newcommand{\bsigi}{\ensuremath{\bm{\Sigma}_i}}

\newcommand{\rij}{\ensuremath{\tilde R_{ij}}}
\newcommand{\raa}{\ensuremath{\tilde R_{11}}}
\newcommand{\rbb}{\ensuremath{\tilde R_{22}}}
\newcommand{\vq}{\ensuremath{\bm{q}}}
\newcommand{\vtheta}{\ensuremath{\bm{\theta}}}

\newcommand{\cj}[1]{\textcolor{red}{(MAX: #1)}}

\begin{document}
\begin{frontmatter}
\title{Leveraging Bayesian Analysis To Improve Accuracy of
  Approximate Models}
\author[add1]{B.T. Nadiga\corref{cor}}
\ead{balu@lanl.gov}
\cortext[cor]{Corresponding author}
\author[add2]{C. Jiang}
\author[add1]{D. Livescu}

\address[add1]{Los Alamos National Laboratory, Los Alamos, NM, USA}
\address[add2]{University of California, Berkeley, CA, USA}

\begin{abstract}
  We focus on improving the accuracy of an approximate model of
  a multiscale dynamical system that uses a set of
  parameter-dependent terms to account for the effects of unresolved
  or neglected dynamics on resolved scales.  We start by considering
  various methods of calibrating and analyzing such a model given {\em
    a few} well-resolved simulations.  After presenting results for
  various point estimates and discussing some of their shortcomings,
  we demonstrate (a) the potential of hierarchical Bayesian analysis
  to uncover previously unanticipated physical dependencies in the
  approximate model, and (b) how such insights can then be used to
  improve the model. In effect parametric dependencies found from the
  Bayesian analysis are used to improve structural aspects of the
  model. While we choose to illustrate the procedure in the
  context of a closure model for buoyancy-driven, variable-density
  turbulence, the statistical nature of the approach makes it more
  generally applicable. Towards addressing issues of increased
  computational cost associated with the procedure, we demonstrate the
  use of a neural network based surrogate in accelerating the
  posterior sampling process and point to recent developments in
  variational inference as an alternative methodology for greatly
  mitigating such costs.  We conclude by suggesting that modern
  validation and uncertainty quantification techniques such as the
  ones we consider have a valuable role to play in the development and
  improvement of approximate models.
\end{abstract}

\begin{keyword}
  Bayesian analysis, reduced order modeling, turbulence modeling,
  Reynolds-averaged Navier Stokes, neural network, surrogate modeling
\end{keyword}

 \end{frontmatter}

 \section{Introduction}
 Natural and engineered systems that exhibit multi-scale behavior due
 to coupling of subsystems with different spatial and temporal scales
 are commonplace. For example, interactions between wave modes,
 interactions between vortical modes, and coupled interactions between wave
 and vortical modes gives rise to complex phenomena in an array of
 fluid dynamic problems ranging from homogeneous isotropic turbulence,
 to common and engineering instances of turbulent fluid flow, to
 rotating-stratified turbulence relevant to flows on global and
 astrophysical scales. In a different setting, interactions between
 fast vibrations and slow conformational changes lead to complex
 behavior in molecular dynamics relevant to protein folding, and so
 on.

 If we represent the comprehensive mathematical
 model of such a complex, multiscale, dynamical system, (the full
 Navier-Stokes equations and the full force-field based molecular
 dynamics respectively in the two examples above) 
 symbolically as
 \begin{equation}
  \frac{d}{dt} \ufo = \ffo(\ufo; \bm p),
  \label{eq:symbolicFO}
\end{equation}
where $\vv u$ represents the state vector, $\vv f$ the tendency of the
state vector, $\vv p$ is a set of physical parameters, and the
superscript $W\!R$ stands for well-resolved, then it is almost
invariably the case that the computational complexity of such a model
prevents it from being used routinely in design and analysis of the
system of interest. Therefore, a pragmatic consideration in seeking an
approximate model is to reduce the computational cost of the orginal
comprehensive model by limiting the associated number of degrees of
freedom, while keeping the error in relevant quantities of interest
(QoI) small. In the context of (\ref{eq:symbolicFO}), an approximate
model may be symbolically represented as
 \begin{equation}
  \frac{d}{dt} \uro = \fro(\uro; \bm p; \bm\phi),
  \label{eq:symbolicRO}
\end{equation}
where superscript $RO$ stands for reduced-order, dimension of
$\uro \ll$ dimension of $\ufo$, and $\bm\phi$ is a set of modeling
parameters.

We note that a Reduced Order Model (ROM) is one kind of an approximate
model that is popular in the engineering fields and industry wherein
the original governing equations are projected on to a reduced-order
basis, but where the basis vectors are determined from snapshot data,
e.g., using Principal Component Analysis (PCA) \cite[e.g., see][and
references therein]{jolliffe2011principal}, or Dynamic Mode
Decomposition (DMD) \cite[e.g., see][and others]{schmid2010dynamic},
etc., and where the snapshot data is obtained by solving the original
system.  In the ROM context, the model parameters $\bm\phi$ may be
thought of as associated with terms such as closures \cite[as in,
e.g.,][and others]{wang2012proper}, etc.. We also note that an
approximate model may be based entirely on a data-driven approach. For
example, in the field of molecular dynamics (MD), it is common to use
an approximate approach such as a Markov State Model (MSM) to permit
longer simulation times. In this approach the conformation space is
first partitioned, based on MD simulation data, into discrete
subspaces, by using techniques such as clustering or PCA. And then, in
such a reduced subspace, molecular kinetics itself is reduced to
transitions between the identified discrete subspaces that are
governed by transition probabilities that may be parameterized and
learnt, again, from MD simulation data \cite[e.g., see][and references
therein]{deuflhard2000identification,pande2010everything,
  schutte2011markov, gerber2017toward}.

 \subsection{Approximate Models of Turbulent Flows}
 While further combinations of such data-driven and equation-based
 techniques are clearly possible \cite[e.g.,
 see][]{dorrestijn2013stochastic}, we shift attention to the specific
 context of a turbulent flow. A common objective in the study of such
 a flow, is to be able to accurately compute QoIs of practical
 relevance. However, in a turbulent flow not only are the relevant
 fields (velocity, density, and others) three-dimensional,
 time-dependent and random, but there is a large range of time and
 length scales. Whereas the domain size ($L$) and geometry directly
 affect the flow, the Kolmogorov length---a measure of the smallest
 scales of turbulence---scales as Re$^{-3/4}$ and the Kolmogorov time
 scales as Re$^{-1/2}$ \cite{pope2001turbulent}. Here, Re is the
 Reynolds number, a non-dimensional number that characterizes the
 relative effects of nonlinearity and viscosity, and it is large in
 turbulent flows.

 Thus, if all the scales were to be resolved, as in the Direct Numerical
 Simulation (DNS) procedure, the computational cost would scale as
 Re$^3$. Here, computational cost is estimated as scaling with the
 number of operations $N_{ops} = N_{cells}\times N_{timesteps}$, with
 $N_{cells}\sim (L/\Delta x)^3 \sim \left(Re^{3/4}\right)^3$, where the
 cube accounts for three-dimensionality of the flow and
 $N_{timesteps}\sim Re^{3/4}$ rather than $Re^{1/2}$ since DNS
 practitioners commonly use explicit time-stepping with
 Courant-Frederick-Levy number of O(1). In this case, the {\em
   computational\/} timestep goes down as $Re^{-3/4}$. Such steep
 scaling of computational cost with Reynolds number for a fully
 resolved simulation makes it feasible only for low to moderate
 Reynolds numbers, meaning to say prohibitively expensive for realistic
 flows of interest.

 Computational intractability of being able to fully resolve a
 (high Reynolds number) turbulent flow, leads naturally to the
 question of modeling only a limited range of scales that are of
 direct interest. In this context, the procedure of averaging the
 governing equations over a scale of the order of the smallest scale of
 interest leads to the closure problem. Here by {\em closure} we refer
 to how the neglected degrees of freedom affect the evolution of the
 resolved degrees of freedom.

 In the Large Eddy Simulation (LES) approach \cite[e.g.,
 see][]{pope2001turbulent}, equations are solved for filtered fields
 (with a filter scale $\Delta$) that are representative of the
 larger-scale turbulent motions and these equations include a
 (closure) model for how the unresolved scales affect the resolved
 scales. While the computational cost of this scale-restricted model
 is seen straightforwardly to scale as $\left(L/\Delta\right)^4$, the
 requirement of $\Delta$ to lie within the inertial range of
 turbulence (a range of scales within which dynamics is self-similar)
 renders its computational cost still too high for realistic flows
 given current computational resources.

 For these reasons, in order to obtain a model that attains acceptable
 levels of accuracy while being computationally affordable, an ensemble-average
 (Reynolds-average) of the original Navier-Stokes equations
 (Reynolds-Averaged Navier Stokes or RANS equations) is considered.
 Since the RANS equations solve only for
 the ensemble-averaged fields, a solution procedure using these equations
 is not required to resolve either the inertial range (cf. LES) or the
 dissipative range (cf. DNS). The RANS equations are therefore
 amenable to being solved using much coarser resolutions. The flip
 side of solving the RANS equations at coarse resolutions
 is that in order to attain an acceptable level of accuracy, the
 effects of a large set of unrepresented scales (on resolved scales)
 have to be explicitly modeled. Such closure issues notwithstanding,
 the RANS approach has emerged as the method of choice
 \cite[e.g., see][]{pope2001turbulent} in both understanding realistic
 turbulent flows and to address practical issues of design,
 optimization, and operations in the context of such flows. The RANS
 approach may be written symbolically as
 \begin{equation}
  \frac{d}{dt} \uro = \fro(\uro; \bm p; \bm\phi) = \ffo(\uro; \bm p) +
  \vv m(\uro; \bm\theta),
  \label{eq:symbolicRANS}
\end{equation}
where $\uro$ is the ensemble (or Reynolds) average of $\ufo$
(=$\overline{\ufo}$), so that if $\vv u' = \ufo - \uro$, $\overline{\vv
  u'}$ = 0, $\vv m$ represent the closure terms, and for convenience the
vector of modeling parameters in the RANS approach is represented by
$\bm\theta$.

The modeling of closures in the LES and RANS approaches to
modeling turbulent flows is central to the accuracy
and efficiency of these methods. Such modeling is based largely on
phenomenological understanding of prototypical turbulent flows by
subject matter experts and is mostly deterministic. However,
stochastic methods that typically use stochastic models of turbulent
fluid flow to evolve probability density functions of turbulent fields
such as velocity have also been used to address the closure problem
\cite[e.g., see][]{pope2001turbulent}, and data-driven stochastic
closures are also being considered \cite[e.g., see][and
others]{duan2007stochastic, nadiga2008orientation,
  plant2008stochastic}. In either case, however, improvement of a
model, in terms of reducing model-form related error has almost exclusively been
the domain of subject matter experts and theoreticians. Thus, whereas
a computational approach for such model improvement currently does not
exist, we present one.

 \subsection{Structural and Parametric Uncertainty in the Approximate Model}
 When additional explicit model terms, such as $\vv m$ in
 (\ref{eq:symbolicRANS}), are used to achieve a simplified
 approximation of the full system, the form of such a closure is a
 choice that is made and is, therefore, not unique. It also follows
 that the OoIs in the approximate model depend on the choice of the
 form of the closure. We refer to this as structural or model-form
 uncertainty/error. After the choice of the form of the closure has
 been made, the values of the parameters $\bm\theta$ influence the
 accuracy of the approximate model, leading to parameter
 uncertainty/error.  Even when additional explicit model terms, such
 as $\vv m$ in (\ref{eq:symbolicRANS}), are not used, it is easy to
 imagine hypothetically introducing such tendencies as a means of
 reducing the discrepency between the average of $\ufo$ and
 $\uro$. Indeed such terms are routine when neural networks are used
 to improve the model, but with the crucial difference that the
 dimension of $\bm\phi$ $\gg$ the dimension of $\bm\theta$ in such a
 case (and consequently structural uncertainty is less important).
 Once the choice of the form of the closure has been made, we next
 consider the process of its validation and calibration given
 parametric uncertainty.

\subsection{Model Calibration and Validation}
The calibration and validation of a model, e.g., as in the RANS
approach to modeling turbulent flows is based either on experimental
data or observations or data from numerical studies that resolve
dynamics and physics of the full range of relevant scales. In the case
of some idealized turbulent flows, such calibration data may come from
DNS of the flow. Here, by calibration, we refer to the assignment or
adjustment of values of model parameters, $\bm\theta$ in
(\ref{eq:symbolicRANS}), so as to bring model prediction of QoIs,
$\qro\equiv\qro(\uro)$, in certain scenarios into agreement with known
values of QoIs, $\qfo$, for those scenarios. And, validation refers to
the process of determining the degree to which the model,
(\ref{eq:symbolicRANS}), is an accurate representation of the real
world, here (\ref{eq:symbolicFO}), from the perspective of the
intended uses of the model \citep{roache1998verification,
  babuska2004verification, farrell2015bayesian}.

Different approaches to calibration and validation exist. 
However, calibration and validation of RANS turbulence models has traditionally
relied on point estimates---that is one optimal set of parameters are
sought that best fits the calibration data \citep[e.g.,
see][]{pope2001turbulent, schwarzkopf2016two}:
\begin{equation}
  \bm\theta_{opt} = \argmin_{\bm\theta}||\qfo(\ufo) - \qro(\uro;\bm\theta||.
  \end{equation}
However, it is not
guaranteed that such calibration is optimal. For example, there may be
multiple, possibly very different multi-way balances of phenomena that
lead to similar evolutions of QoIs. In the turbulent flow
context, different balances of dissipation, transport, and decay
processes could be consistent with the evolution of turbulent moments
as specified by given calibration data.

In contrast to approaches that aim to find a single optimal set of
parameters, a Bayesian framework models the model parameters as random
variables and seeks to approximate the joint distribution of such
variables. In other words, it estimates the posterior probability of
$\bm\theta$ given $\qfo$: $P(\bm\theta | \qfo)$. In so doing, the
methodology gives consideration to the possibility that different
balances of key modeled processes can explain the calibrating
data. That is, the Bayesian framework integrally permits consideration
of parameter uncertainty in the calibration and validation of a
model. Furthermore, the Bayesian approach allows for better integration
of both prior knowledge and probabilistic structure into the
calibration and validation process. We are, therefore, interested in
examining the utility of a Bayesian approach to calibration and
validation, and what advantages it may hold over the traditional point
estimate based approaches in developing insights into improving the
model under consideration.

A few recent studies have used Bayesian estimation
techniques to calibrate RANS models. For example,
\citet{oliver2011bayesian}, and \citet{cheung2011bayesian} use
Bayesian uncertainty analysis to calibrate and inter-compare four
well-known RANS models: the Baldwin-Lomax model
\cite{baldwin1978thin}, the SA model \cite{spalart1992one}, the Chien
$k$-$\epsilon$ model \cite{chien1982predictions}, and the $v^2-f$
model \cite{durbin1991near}; \citet{edeling2014bayesian,
  edeling2014predictive} used Bayesian estimates of parameter
variability in the Launder-Sharma $k-\epsilon$ model
\cite{launder1974application} as a means to estimate errors in RANS
simulations; \cite{farrell2015bayesian} present an adaptive modeling
algorithm for selection and validation of models,
however, in the domain of atomistic systems.
In other work that uses a Bayesian framework in the context of RANS
models, \citet{edeling2014bayesian, edeling2014predictive}, note that
the distribution of  model parameters provides information on error
associated with the  model: when the joint probability density function
(PDF) of the  model parameters is propagated through the  model, the
distribution of the QoIs can be used to provide confidence bounds for
the QoIs. 

In addition to such uses of the Bayesian methodology, we think that
the Bayesian methodology has a useful role to play in the development
and improvement of the approximate or reduced-order descriptions---a role
that should be thought of as complementary to that of subject matter
expertise that remains central to developing approximate models. Indeed, we show
how Bayesian analysis of a RANS model that uses DNS (or equivalently
any other calibration data) can uncover unanticipated dependencies
which in turn point to structural deficiencies in the model and
specific ways in which the model can be improved.
%

Organization of the rest of the article is as follows. In
Sec. \ref{sec:turbmodel} we discuss the specifics of the problem we
consider, the fully resolved model, the specific approximate
model we consider, and the QoIs. In Sec. \ref{sec:methods} we discuss
the details about the methods we use before presenting respective
results. This includes details about point estimates and Bayesian
estimation, how we go about improving the model and
details about the neural network-surrogate strategy that we propose
for situations wherein the approximate model itself may still
be expensive enough to prohibit its direct use in Bayesian
analysis. This is followed by a brief section that summarizes the work
and concludes.

\section{The Problem Setting, the Fully Resolved Model, and the
  Approximate Model}\label{sec:turbmodel}

Buoyancy-driven turbulence is commonplace in a wide variety of
naturally-occuring flows (e.g., ocean-atmosphere dynamics,
astrophysics, mantle convection etc.) and engineering flows (e.g.,
ranging from smoke-stacks to combustion to inertial-confinement
fusion), and encompassing both, a wide range of instabilities and rich
phenomenology. Homogeneous Rayleigh-Taylor (hRT) turbulence is a particular
idealization of buoyancy-driven turbulence.
The initial condition for the onset of hRT turbulence consists of
isolated regions in the domain of interest being occupied by two
miscible fluids at rest that have different densities, and the flow
evolves in accordance with the resultant buoyancy force. 

\subsection{The Fully Resolved Model}
If the flow is incompressible (low-speed) and the densities of the two
fluids are $\rho_1$ and $\rho_2$, the full-order governing equations
(symbolically represented by
Eqn. \ref{eq:symbolicFO}) for this problem are the Navier-Stokes
equations along with the species mass
fraction transport equations.

\paragraph{Species mass fraction transport equations} Given mass fractions $Y_1 = Y_1(\bm{x}, t)$ and $Y_2 = Y_2(\bm{x}, t)$ for the two fluids which sum to unity (i.e., $Y_1 + Y_2 = 1$) , the density of the mixture can be written as:
\begin{align}
    \rho = \frac{1}{Y_1 / \rho_1 + Y_2 / \rho_2}
\end{align}
The species mass fraction transport equations assuming Fickian diffusion are:
\begin{align}
    (\rho Y_{\alpha})_{,t} + (\rho Y_{\alpha}u_j)_{,j} = (\rho D_0 Y_{\alpha_{,j}})_{,j}.
\end{align}
The continuity equations is obtained by summing over $\alpha = 1, 2$. Non-zero divergence of velocity results from mixing due to the change in specific volume, $1/\rho$:
\begin{align}
    u_{j,j} = -D_0 \ln{\rho_{,jj}}.
\end{align}
Here $D_0$ is the diffusion coefficient and it is assumed to be constant. 

\paragraph{Naver-Stokes equations} After non-dimensionalization, using an arithmetic
mean of the two densities for the reference density
$\rho_0 = \frac{1}{2}(\rho_1 + \rho_2)$, a reference velocity $U_0$,
and a reference length $L_0$, the well-resolved model equations are:
\begin{align}
    \rho_{,t} + (\rho u_{j})_{,j} &= 0, \\
    (\rho u_i)_{,t} + (\rho u_i u_j)_{,j} &= -p_{,i} +\tau_{ij,j} + \frac{1}{Fr^2}\rho g_i, \\
    u_{j,j} &= -\frac{1}{Re_0 Sc}\ln{\rho_{,jj}},
\end{align}
and where
$\tau_{ij} = (1/Re_0)(u_{i,j} + u_{j,i} - (2/3)u_{k,k}
\delta_{ij})$.  In the above equations, the non-dimensional state variables
are the density $\rho$, $x_i$-direction velocity $u_i$, and
pressure $p$. The 
non-dimensional parameters in the above equations are the Reynolds number $Re_0$,
Schmidt number $Sc$, and Froude number $Fr$, defined by:
\begin{align}
    Re_0 = \frac{\rho_0 L_0 U_0}{\mu_0}, \quad
    Sc = \frac{\mu_0}{\rho_0 D_0}, \quad
    Fr^2 = \frac{U_0^2}{gL_0}
\end{align}
and where $g$ is gravitational acceleration, and $\mu$ is for dynamic
viscosity (assumed constant and equal for both fluids). In the above equations, we note that (a) because we consider large
density differences, the flow is not amenable to the commonly used
Boussinesq approximation, leading to the full density being used
consistently in all the terms of the momentum equations, and (b) even
though we consider low-speed dynamically-incompressible flows, the
velocity field is not divergence-free because molecular mixing leads
to changes in the specific volume.

Equations (2.8)-(2.10) are solved using the Direct Numerical Simulation
procedure at four different values of Atwood number: 0.05, 0.25, 0.50,
and 0.75, where the Atwood number is the normalized density ratio. The reader is referred to
\cite{livescu2007buoyancy} for details. These values of the Atwood number
characterize the initial conditions, when the fluids are completely segregated. 
As the fluids molecularly mix, the evolving Atwood number can be calculated as 
the largest value of the (normalized) density difference between pairs
of points in the flow.
Such an evolving Atwood number decreases with time.

A brief phenomenological description of the flow evolution is as
follows, and the reader is referred to \cite{livescu2007buoyancy} for
further details. The unstable nature of the initial distribution of
density endows it with high potential energy, and the ensuing
buoyancy-driven instability leads to a conversion of potential energy
into kinetic energy. The flow subsequently transitions to turbulence. Once 
turbulent, the flow can be described in terms of the evolution of
single-point second-order turbulent correlations. The relevant
turbulent correlations in this setting are the Favre average Reynolds
stresses $\tilde R_{ij}$ ($= \overline{\rho u_i''u_j''}/\bar{\rho}$,
 where $u_i''= u_i - \tilde u_i$, with $\tilde u_i$ being the Favre or
density-weighted averaged velocity $\overline {\rho u_i} / \bar\rho$)
and the turbulent kinetic energy ($K \equiv \tilde R_{ii}/2$), the turbulent
mass flux $a_i$ which is the correlation between density 
and velocity fluctuations
($a_i\equiv\bar u_i''=\overline{\rho ' u_i'}/\bar{\rho}$), and the
density-specific-volume covariance $b$.  The density-specific volume
correlation is largest at initial times and begins to decay as the
turbulence grows and the fluids mix (see solid
lines in Fig.~\ref{fig:pointestimate-qois}). In particular, at these initial
times, the mean pressure gradient that develops in response to gravity
couples to $b$, to produce turbulent mass flux $a_1$, which then
couples back with the pressure gradient to generate turbulent kinetic
energy. At early to intermediate times, the Reynolds stresses (and turbulent
kinetic energy), the turbulent mass flux, and turbulent dissipation all grow.
Eventually, as the fluids mix, they reach peak values at different
times and eventually decay asymptotically to zero.

In this problem setting, the QoIs are the various second order
turbulent correlations described previously, and a length scale as
discussed further in the next section.  Temporal evolution of some of
the QoIs, as given by the fully resolved model, is shown in solid
lines in Fig.~\ref{fig:pointestimate-qois}.

\subsection{The Approximate Model} 
In this context, and in continuation of the long
history of turbulence model development,
\citet{schwarzkopf2011application} showed that the single point
turbulence equations developed by \citet{besnard1992turbulence} (BHR
model) could be applied to a range of self-similar turbulent mixing
flows generated from different instabilities such as Rayleigh-Taylor,
Kelvin-Helmholtz, and Richtmyer-Meshkov without changing the model
coefficients. Key to the wide applicability of the model developed by
\citet{besnard1992turbulence} was their consideration of a transport
equation for density specific volume covariance $b$. This, in conjunction
with full consideration of the Reynolds stress transport, allowed the
proper description of anisotropy that is fundamental to
buoyancy-driven turbulence. Nevertheless, a shortcoming of this model
was its inability to properly represent turbulence growth rates
encountered in settings that involve transient evolution of
buoyancy-driven turbulence, as commonly encountered in flows
dominated by the Raleigh-Taylor instability. This shortcoming is
better understood by considering the popular $k$-$\epsilon$ class of
turbulence models wherein the transport term is slaved to the decay
length scale (by coefficients $C_\mu$ and $\sigma_k$). Consequently,
like in the $k$-$\epsilon$ model, profiles of dissipation and
transport of Reynolds stress (and density-specific volume covariance)
are scaled versions of each other (in the BHR model). This is in
contrast to the different nature of the scaling of transport and
dissipation that is exhibited in the self-similar regime of
Rayleigh-Taylor turbulence in \cite{livescu2009high}. To remedy this
shortcoming, \citet{schwarzkopf2016two} introduced a second turbulent
length scale.

The RANS turbulence model discussed above contains several
coefficients. These coefficients need to be calibrated so that a
common set of model coefficients will allow reasonable comparisons of
statistics over a wide range of turbulent flows,
ranging from incompressible flows with single fluids and mixtures of
different density fluids (variable density flows) to flows over shock
waves. \citet{schwarzkopf2016two} follow a recipe for calibrating the
coefficients that considers a sequence of simple flow
configurations. The simple flow configurations considered were such
that most of the terms in the equations could be neglected and the
remaining non-zero terms were (mostly) different for each of the configurations
considered. The specific sequence of canonical flows considered in the
calibration process consisted of homogeneous isotropic decaying
turbulence, homogeneous buoyancy-driven decaying turbulence,
homogeneous shear (including Rapid Distortion Theory at high shear rates),
wall bounded flow, Rayleigh-Taylor (RT) driven turbulence, shear driven turbulence, and shocked isotropic
turbulence.

A RANS turbulence model simultaneously represents a number of physical
processes (e.g., production dissipation, return to isotropy and rapid
distortion of second order turbulent velocity correlations and
others). Therefore, it seems important that the calibration process
should account for the possibility that different combinations of such
processes can lead to QoIs, either as observed in experiments or as
computed in DNS. This amounts to requiring a comprehensive
consideration of uncertainty in the coefficients (parametric
uncertainty). The traditional calibration process adopted in
\cite{schwarzkopf2016two}, by producing one optimal combination of
coefficient values, fails to account for such parametric
uncertainty. In this sense, it is clear that a calibration process
that accounts for such uncertainty will be better than the calibration
process adopted in \cite{schwarzkopf2016two}. While we are currently
working on a comprehensive Bayesian calibration of the RANS turbulence
model that includes a full complement of test cases (e.g., as
considered in \cite{schwarzkopf2016two}) that will not be the focus of
the current article. The focus of this study, instead, is on
demonstrating how the diagnostics resulting from the Bayesian
calibration and analysis can be leveraged to improve turbulence
modeling. For this, it suffices to consider a suite of homogeneous
buoyancy-driven or Rayleigh-Taylor (hRT) turbulence test cases.

The RANS turbulence model equations for hRT turbulence is a set of
ordinary differential equations:

\begin{alignat}{2}
&\frac{d \raa}{d  t} &&= \Big(2-\frac{4}{3}C_{r1}\Big)a_1\frac{\bar{P}_{,1}}{\bar{\rho}}-C_{r3}\frac{\sqrt{K}}{S_D}\Big(\raa - \frac{2}{3}K\Big)-\frac{2}{3}\frac{K^{3/2}}{S_D}\label{eq:ge1}\\
&\frac{d \rbb}{d  t} &&= \frac{2}{3}C_{r1}a_1\frac{\bar{P}_{,1}}{\bar{\rho}}-C_{r3}\frac{\sqrt{K}}{S_D}\Big(\rbb -\frac{2}{3}K\Big)-\frac{2}{3}\frac{K^{3/2}}{S_D}\label{eq:ge2}\\
&\frac{d a_1}{d  t} &&= (1-C_{ap})\frac{b}{\bar{\rho}}\bar{P}_{,1}-C_{a1}\frac{\sqrt{K}}{S_D}a_1\label{eq:ge3}\\
&\frac{d b} {d  t} &&= -C_{b1}\frac{\sqrt{K}}{S_D}b\label{eq:ge4}\\
&\frac{d S_D}{d  t} &&= \frac{S_D}{\bar{\rho}K}\Big(\frac{3}{2}-C_{4v}\Big)a_1\bar{P}_{,1}-\Big(\frac{3}{2}-C_{2v}\Big)\sqrt{K}.\label{eq:ge5}
\end{alignat}
Here $t$ is time, $\bar\rho$ is the mean material density,
$\bar{P_{,1}}$ is the mean pressure gradient in the direction of
gravity, $\tilde R_{ij}$, $K$, $a_i$, and $b$ are the Favre average Reynolds stress, turbulent
kinetic energy, mean density-weighted velocity fluctuation, and density-specific-volume,
respectively, as defined previously. $S_D$ is the turbulent decay length scale. The remaining
terms denoted by subscripted $C$s are model parameters and consist of $C_{r1}$, related
to  rapid return to isotropy; $C_{r3}$ related to  slow return to
    isotropy; $C_{ap}$, related to rapid decay of mass flux; $C_{a1}$,
    related to slow decay of mass flux;
    $C_{4v}$, related to rapid growth/decay for hRT; and $C_{b1}$,
    related to decay of
    density-specific volume. $\raa, \rbb, a_1, b$ and
turbulent kinetic energy $K$ are the QoIs resulting
from this model. We refer readers to \citet{schwarzkopf2016two} for
further details about the RANS turbulence model. For convenience, this set
of equations may be written as
\begin{equation}
  \frac{d}{dt} \TRANS = \vv F(\TRANS, \uro; \bm \theta),
  \label{eq:symbolicqrans}
\end{equation}
where $\vv q$ is the vector of turbulent second-order moments of
interest of the primitive variables $\vv u$, and $\bm\theta$ is the set of
parameters. In this setting, the closure term $\vv m(\uro; \bm\theta)$
in (\ref{eq:symbolicRANS}) is given by the turbulence model:
\begin{equation}
  \vv m(\uro; \bm\theta) = \vv m(\TRANS(\uro; \bm\theta))
  \label{eq:mq}
\end{equation}
Here it is also understood that the RANS turbulence
model is an approximation of 
\begin{equation}
    \frac{d}{dt} \TDNS = \FDNS(\TDNS, \ufo),
\label{eq:qexact}
\end{equation}
but where $\FDNS$ is not explicitly known because of the dependence of
$\FDNS$ on yet higher order moments and so on. As such, the closures
introduced in (\ref{eq:symbolicRANS}) to approximate
(\ref{eq:symbolicFO}), equivalently approximations introduced in
(2.17) to approximate (2.19), invariably lead to model error that may be
attributed to structural inadequacy/error and parametric errors as
discussed in the introduction. While we are interested in improving
the accuracy of (\ref{eq:symbolicRANS}) by improving $\vv m$, since
$\vv m$ is determined from $\TRANS$ as in (\ref{eq:mq}), in the rest
of this article, unless otherwise stated, we refer to the RANS
turbulence model given by the set of equations (\ref{eq:ge1}) to
(\ref{eq:ge5}) or equivalently (\ref{eq:symbolicqrans}) as the approximate model we
consider and wish to improve.

\section{Methods and Results}\label{sec:methods}
There are multiple ways in which the set of parameters in
(\ref{eq:ge1}-\ref{eq:ge5}) can be estimated given a set of data
such as data from a few DNSs. For convenience, we broadly categorize
the estimation methods based on two considerations: first, based on
whether the estimation procedure pools all of the sample data together
or not (see below), and, secondly, based on whether the method produces a
point-estimate or a probabilistic or interval estimate.
Since the distinction between methods based on the second
consideration is evident, we do not devote a separate section to point
them out; rather we discuss relevant details in the respective sections
that present the results. 

\subsection{Pooled and Unpooled Analysis}\label{sec:pooled-unpooled}
To motivate the distinction between pooled and unpooled analysis, we
note that we are interested in analyzing the RANS turbulence model
given DNS simulations that have been previously performed at four
different Atwood numbers: $\text{At} \in\{0.05, 0.25, 0.50, 0.75\}$.
The aim in performing such an analysis is to then examine the
resulting point-estimates or posterior distribution of parameters in
the turbulence model as a means to gain insights into the turbulence model itself. For example, if an {\em a priori} unexpected
dependency is found in the study, it would lead us to further
investigate the origin of the dependency in terms of turbulence
phenomenology. If this investigation leads us to conclude that the
dependency is a shortcoming of the model, further follow-up would
consist of finding appropriate (possibly structural) modifications of
the turbulence model that will remove the dependency while not
introducing yet other ``unexpected/unwanted'' dependencies.

To this end, we note that the sample data is at different Atwood
numbers, a parameter that characterizes the strength of the drive that
is forcing the turbulent flow. At the same time, the RANS turbulence
model in (\ref{eq:ge1}-\ref{eq:ge5}) does not explicitly depend on
Atwood number. That is because of the local nature of the differential
equation form of the RANS turbulence model; the RANS turbulence model
subsumes such dependencies through its dependence on local turbulent
correlations. That is, the RANS approach aims to model the flow
evolution based purely on local behavior of density and related
quantities. As such, assuming that the given turbulence model is
correct leads us to consider the sample data as coming from the same
population, and to estimate a single set of parameters for the
different Atwood numbers. We call this the pooled analysis. A
comparison of the QoIs with this point estimate would then allow us to
estimate the effectiveness of the turbulence model.

If the RANS estimates of the QoIs do not compare well to the DNS
estimates, two possibilities exist: either the sample data did not
come from the same population (still implying a deficiency in the
turbulence model) or the turbulence model itself is (more seriously)
deficient.  In order to examine the possibility that the sample data
do not come from the same population, we need to perform an
``unpooled'' analysis. That is, we need to assume that the data at
different Atwood numbers are coming from different populations, and
redo the analyses to obtain a separate estimate for the set of
turbulence parameters at each Atwood number. We call this the unpooled
analysis. We note that while indeed partial pooling is also possible
and useful in certain circumstances, it suffices to consider the two
end members, viz. fully pooled or pooled and unpooled analysis, for
our present purpose.

If  the QoIs compare well with DNS with the unpooled analysis, then
the dependence of the parameters on Atwood number would have to be
investigated further to be able to remove it from the
turbulence model. Needless to mention, a more serious shortcoming of
the turbulence model would be implicated in the case of a poor
comparison between the RANS and DNS estimates of QoIs on performing
unpooled analysis.

\begin{algorithm}[!t]
  \SetAlgoLined \textbf{Input:} initial parameters $\vtheta$, RANS estimate of QoIs as a function of
  $\vtheta$, $\qrans(\vtheta)$, DNS QoI
  $\qdns$, step size $\lambda$\, accuracy tolerance $\tau$\;
  \textbf{Output:} Point estimation of optimal parameter
  $\tilde{\vtheta}$.\; \For{$i\gets0$ \KwTo $N_{step}$}{
    $\epsilon = ||\qrans-\qdns||_2$\;
    $\Delta \vtheta = - \lambda \nabla_{\vtheta}\epsilon$\;
    \If{$\Delta \vtheta < \tau \vtheta$}{\textbf{break for}\;}
    $\vtheta\gets \vtheta + \Delta \vtheta$\; }
  $\tilde{\vtheta}\gets \vtheta$
\caption{Gradient-based point estimation algorithm. The gradient
  $\nabla_{\vtheta}$ can be acquired using either forward
  sensitivities from finite difference, or adjoint sensitivities. For
  pooled estimates, the residuals are summed (in the L2 sense) over the Atwood numbers
  as well, whereas for the unpooled estimates, the procedure is
  conducted separately for each Atwood number.}\label{alg:pt}
~
\end{algorithm}

\newcommand{\fgprfx}{Figs/}
\begin{figure}
\bc
\includegraphics[width=0.9\textwidth]{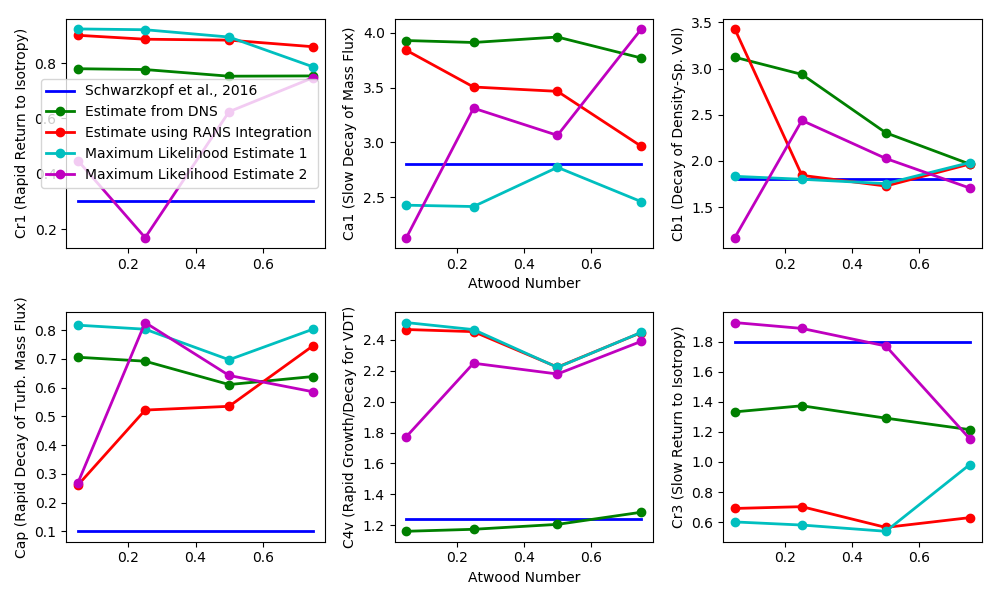}
\ec
\caption{Results from calibration of the RANS model using point
  estimates. The residual between the DNS and RANS estimates of QoIs
  (objective function) is minimized. Not only are the estimates of the
  various parameters different, but their variation with Atwood number
  is also different.}
\label{fig:pointestimate}
\end{figure}

\begin{figure}
\bc
\includegraphics[width=0.49\textwidth]{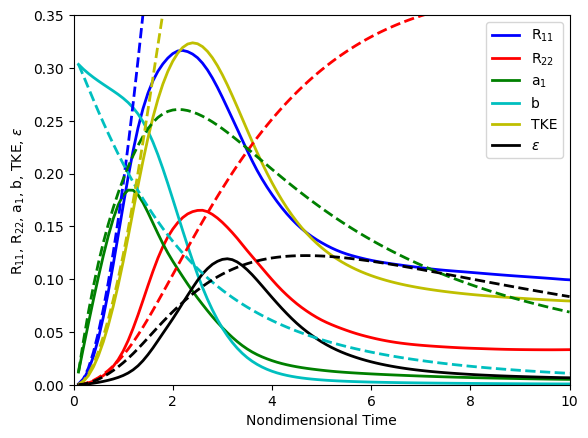}
\includegraphics[width=0.49\textwidth]{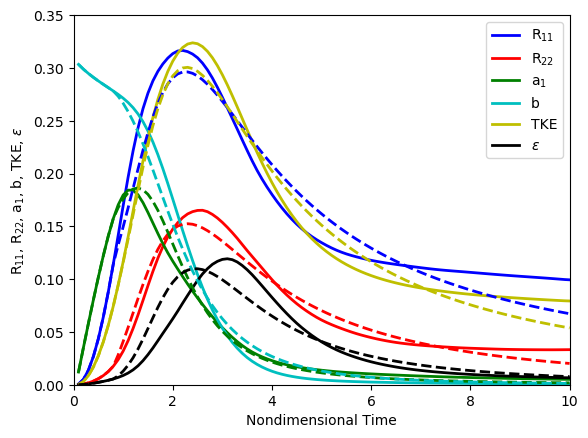}
\ec
\caption{QoIs using two of the point estimate-based
  calibrations. Solid lines are from DNS. Dashed lines are RANS
  estimates of the QoIs. The panel on the left uses the parameter
  settings of \cite{schwarzkopf2016two}. The panel on the right uses
  nonlinear least-squares to find the optimal set of parameters for
  each of the Atwood number cases separately (unpooled). In Table~1,
  the left panel corresponds to row 1, and the poor comparison of QoIs
  is indicated by larger residuals. The right panel corresponds to
  row 3, and the better comparison of QoIs is indicated by the reduced
  residuals. Atwood number is 0.50.}
\label{fig:pointestimate-qois}
\end{figure}

\subsection{Point Estimates and Results}
In this category of estimation methods, the RANS model is used in
conjuction with calibration data to calculate a single set of values
for the turbulence parameters---a set that serves as
the best estimate for those parameters given the data. Figure~1 shows
five different point estimates. The first estimate, in red, corresponds to that
of Schwarzkopf et al., 2016. This may be considered as an example of a
pooled point-estimate. Other pooled point-estimates are not shown to
avoid clutter.

In the second approach all quantities other than the parameters in
(\ref{eq:ge1}-\ref{eq:ge5}) are evaluated from the DNSs to obtain a
set of algebraic equations in the unknown parameters. These equations
are then solved in the least-squares sense to obtain the second, now unpooled,
point-estimate of the parameter set (at each Atwood number for which the
DNSs were available). This is shown in blue.

For the third estimate (green), an optimal set of parameters is sought
at each Atwood number while integrating the RANS model. We note that
we implemented estimation procedures using both forward sensitivities
and adjoint sensitivities of the QoIs with respect to the parameters
$\vv\theta$. Differences in these estimates were not significant, as
expected, while noting that the adjoint sensitivities become more
attractive when the number of parameters is larger than in the present
case.

Next, we consider unpooled ``maximum
likelihood''estimates (MLE). The likelihood function is given by
(\ref{eq:likelihood}) and is described further in the section on
Bayesian analysis. Two different estimates, (in cyan and magenta) are 
considered in this case to highlight another issue with
point-estimates---their dependence on the initial value provided to
the search algorithm. 

\begin{table}
\centering
\begin{tabular}{llcccc}  
\toprule
&Experiment & At=0.05 & At=0.25 &At=0.50 &At=0.75 \\
\midrule
1& Schwarzkopf et al. (pooled) & 0.408 & 1.726 & 0.92 & 2.258\\
2& Unpooled DNS Only & 0.144 & 0.645 & 1.088 & 1.792\\
3& Unpooled RANS Integration & 0.023 & 0.143 & 0.293 & 0.384 \\
4& Unpooled Max. Likelihood 1 & 0.029 & 0.156 & 0.302 & 0.392\\
5& Unpooled Max. Likelihood 2 & 0.069 & 0.166 & 0.385 & 0.476\\
\midrule
6& Pooled Bayesian & 0.060 & 0.224 & 0.678 & 1.596\\
7& Unpooled Bayesian & 0.029 & 0.160 & 0.327 & 0.487\\
8& Pooled Modified Bayesian & 0.028 & 0.169 & 0.500 & 0.439\\
9& Unpooled Modified Bayesian & 0.029 & 0.160 & 0.331 & 0.498\\
10& DNN-Surrogate Bayesian & 0.028 & 0.159 & 0.326 &0.515\\
\bottomrule
\end{tabular}
\caption{The (uniformly-weighted) combined residual at different Atwood numbers for the
  calibrations that are presented. Note that for the maximum
  likelihood estimates and all Bayesian estimates, the likelihood
  function (\ref{eq:likelihood}) is different from the uniformly-weighted combined residual
  presented here. }
\end{table}

The corresponding behavior of the QoIs obtained on
(re)integrating the RANS model using two of the five parameter estimates
discussed above (one pooled and one unpooled) are shown in
Fig. 2. First, the estimate from \cite{schwarzkopf2016two} is seen to result
in poor comparisons of the RANS and DNS estimates of the QoIs. To a
first order, the poor performance of the first estimate could be
attributed to the fact that the  calibration procedure of \cite{schwarzkopf2016two}
considers a number of test-flows and produces one set of parameters to
be used across them. Next, the second (now unpooled) estimate produces
similarly poor comparisons in the QoIs (see Table~1).  The poor performance of the
second estimate can be understood as due to not fully considering the
dynamical nature of the RANS model when estimating the parameter
values \citep[e.g., see][]{nadiga2007instability}. With the third estimate, the
correspondance between RANS and DNS estimates of the QoIs is seen to
be good. This is also true of the  fourth and fifth (both unpooled) estimates;
see Table~1.

The uniformly-weighted combined residuals for the five QoIs at
different Atwood numbers for these five estimates are presented in the
first five rows of Table~1, Here the residuals are the differences
between the DNS and RANS estimates of the QoIs. These residuals are
consistent with the above characterization of the parameter estimates.

The good correspondance between RANS and DNS estimates of QoIs seen in Fig.~\ref{fig:qoi}
with the third, fourth, and fifth estimates establishes that the
turbulence model being considered has the phenomenlogy required to
represent the class of flows being considered.  Furthermore, along the
lines of reasoning for the poor performance of the first (pooled)
estimate, the better performance of the third through fifth estimates may
again be understood as due to the fact that the calibration was
specific to each of the different Atwood number cases (unpooled).

Next, as discussed in Sec.~4.1, the favorable outcome with some of the
unpooled estimates leads us to focus attention on the variation of the
parameter estimates with Atwood number.  Large variations with Atwood
number are seen in most of the six parameters when the third through
fifth estimates are considered. This would suggest that multiple
aspects of the turbulence model have to be modified. However, as good
as the match is between DNS and RANS at these estimates, we note that
this estimation procedure does not account for parametric uncertainty
discussed previously. We therefore turn our attention next to
probabilistic estimates.

\subsection{Bayesian Analysis Methodology}
In the introduction and in the preceding parts of this section, we
briefly considered the role of calibration in the overall process of
validation and discussed how uncertainty has an important role to
play. Here, we outline the Bayesian approach to calibration wherein
uncertainty is quantified in a probabilistic sense. In the Bayesian
framework, calibration of a model is formulated as an inference
problem. Thus, the probability of (the vector of) the model
parameters $\vtheta$, given experimental or DNS data or QoIs $\vq$, is
determined using the Bayes rule as

\begin{align}
  \label{eq:bayes}
  P(\vtheta | \vq) =& \frac{P(\vq | \vtheta)P(\vtheta)}{\int P(\vq|\vtheta') P(\vtheta') d\vtheta'} \\
  \propto& L(\vq | \vtheta) P(\vtheta) \label{eq:bayprob}
\end{align}
In the above equation, it is common to call 
\begin{itemize}
\item 
$P(\vtheta)$ the prior
distribution of the model parameters. 
\item 
$L(\vq | \vtheta)$ the likelihood function, and 
\item 
$P(\vtheta | \vq)$, the posterior distribution of the model parameters
\end{itemize}
An analytical approach to Bayesian
estimation requires the integration of the normalization term
(denominator in (\ref{eq:bayes}). This is difficult to evaluate when
the probability model structure is complex. However, algorithms based on the Monte-Carlo
Markov Chain (MCMC) allow for discretely sampling points from the
posterior distribution, bypassing the need for the integration of the normalization
factor. The only requirement for performing MCMC based sampling is
that a function proportional to the original probability distribution
be specified. This is shown in 
(\ref{eq:bayprob}). To assist in the illustration of the Bayesian
estimation process, we further make precise the QoIs
vector $\vv q$ as follows:
\begin{itemize}
\item $\qdns\in\mathbb{R}^{N_t}$ is the $i$-th QoI
  vector from the DNS simulation data. $i\in\mathbb{Z}, i\in[1, N_q]$
  where $N_q$ is the number of QoIs in the problem. $N_t$ is the total
  number of time-steps. Denote this as the truth vector.
\item $\qrans(\vtheta)\in\mathbb{R}^{N_t}$, results from RANS model,
  given model parameters $\vtheta\in\mathbb{R}^{N_p}$ where $N_p$ is
  number of RANS model parameters. Denote this as the model vector.
\end{itemize}

\paragraph*{Likelihood function and discrepancy model} We parameterize the likelihood
function $L(\vq|\vtheta)$ as a multivariate Gaussian density
function centered around the truth vector. For the $i$-th QoI,
\begin{align}
L_i(\vq|\vtheta) &= f(\qrans(\vtheta);\qdns,\bsigi) \\
&= \frac{1}{(2\pi)^{N_t / 2}|\bsigi|^{1/2}}\exp\Big(-\frac{1}{2}(\qrans-\qdns)^T\bsigi^{-1}(\qrans-\qdns)\Big)
\label{eq:likelihood}\end{align}
Here $\bsigi$ is the covariance matrix (or equivalently $\bsigi^{-1}$
is the precision matrix) for the $i$-th QoI. That is,
when comparing $\qrans$ to $\qdns$, we model the discrepancy between
them that arises due to various reasons including model error as
$\qdns = \qrans + \epsilon_{it}$, with
$\left<\epsilon_{it}\epsilon_{is}\right> = \bsigi$ and where
$\bsigi(t,s)$ represents the assumed covariance structure of the
discrepency. We present results here for a simple parameterization of
$\bsigi(t,s)$ as $\sigma_i \delta(t-s) \equiv \sigma_i \bm{I}$, and
where $\sigma_i$ is a new hyper-parameter and $\bm{I}$ is the identity
matrix. Clearly, while more complicated parameterizations such as
Gaussian processes \citep[e.g., see][]{oliver2011bayesian,
  nadiga2018dependence} and other forms can be used for the
parameterization of the covariance, the simple form for the
discrepancy that we use here, including its independence of $\vtheta$
follows from our {\em a priori} lack of knowledge about such
dependencies including {\em a priori} lack of knowledge of model
error. Finally, a limited amount of experimentation with the more
complicated forms of the covariance showed robustness of the results
we find with respect to the form of covariance. 

Finally, we  define the likelihood function
of all QoIs as a sum of the individual likelihood functions, i.e.,
\begin{align}
L(\vq|\vtheta) = \sum_{i=1}^{N_q}L_i(\vq|\vtheta)
\end{align}
where $\vtheta$ is now augmented to include ${\bm \sigma}$, the $N_q$
dimensional vector of hyper-parameters $\sigma_i$: $\vtheta =
(\vtheta^p, \bm\sigma^q)$.
Finally, given the parameter vector $\vtheta$, $\TRANS$ is obtained by
accurate numerical integration of (\ref{eq:symbolicRANS}).

\paragraph*{Prior distributions} To ensure robustness of the analysis
and results presented, all computations were performed with two
different forms for the prior distribution for parameters $\vtheta^p$,
a normal distribution and a uniform distribution. The priors are
always centered at the previous estimates of
\citet{schwarzkopf2016two}. As mentioned in the introduction, if the
processes associated with the parameters play an important role in the
class of flows we consider, we use weak, uninformative and diffuse
priors; for other parameters, the width of the priors are broadly
determined by physical bounds such as trying to limit regions
of negative turbulent kinetic energy.  Broad normal
distributions are used for $\bm\sigma^q$.


\paragraph*{MCMC using Delayed Rejection Adaptive Metropolis (DRAM)}
We use the DRAM algorithm \cite{haario2006dram}, which is an improved MCMC
algorithm, for drawing samples from the posterior distribution. DRAM
combines two ideas that improve on the Metropolis-Hasting type MCMC
algorithm, Delayed Rejection \cite{tierney1999some} and Adaptive
Metropolis \cite{haario2001adaptive}, whose efficiency in many
scenarios outperform the original methods. The pseudo-code for the
algorithm is given in Algorithm~2, and we limit $N_{stage}$ to two.

\begin{algorithm}[!t]
\SetAlgoLined
\textbf{Input:} Initial point $\vtheta^{(0)}$, initial covariance
$\bm{\Sigma}^{(0)}$,\\ function to compute posterior probability $P(\vtheta | \qfo)$\;
\textbf{Output:} Bayesian estimates for parameters \{$\vtheta^{(i)} | i = 1, 2, \cdots, N_{samp}$\}\;
 \For{$i\gets1$ \KwTo $N_{samp}$}{
    \For{$t\gets 1$ \KwTo $N_{stage}$}{
      $\bm \Theta^{(t)} = \vtheta^{(i-1)} +
    \bm\xi^{(t)},  \quad \bm\xi^{(t)} \sim \bm N(\bm 0, \frac{1}{3^{(t-1)}}\bm{\Sigma}^{(i-1)})$\;
    $\alpha^{(t)}\gets\alpha(\bm\Theta^{(k)},P(\bm\Theta^{(k)} | \vq),
    P(\vtheta^{(i-1)} | \vq), \; k=1 \,\KwTo \, N_{stage})$ \;
    \text{// Compute acceptance probablilty at stage t. e.g.,
    }$\alpha^{(1)} \gets P(\bm\Theta^{(1)} | \vq) / P(\vtheta^{(i-1)}
    | \vq)$ \;
  \If{$u\sim U(0,1)\leq \alpha^{(t)}$}{
  $\vtheta^{(i)}\gets \bm\Theta^{(t)}$ \text{// accept}\;
  \textbf{break for}\;
  }
  \If{$i = N_{samp}$}{
  $\vtheta^{(i)}\gets \vtheta^{(i-1)}$}
}
 \eIf{$i < N_{samp}/2$}{
  $\bm{\Sigma}^{(i)} \gets \eta \,\left(cov(\vtheta^{(0)}, \cdots,
  \vtheta^{(i)}) + \delta \bm I\right), \quad \delta = 10^{-5}$
  \text{// update proposal covariance periodically using entire chain
    up to half of total samples}\;
  \text{// Adapt scaling factor $\eta$ towards 23.4\% acceptance after
  starting at 2.4$^2$/dimension($\vtheta$)}\;  
  }{
  $\bm{\Sigma}^{(i)}\gets \bm{\Sigma}^{(i-1)}$
  }
}
\caption{Bayesian estimation of parameters using the Delayed Rejection
  Adaptive Metropolis (DRAM) variant of the
  random-walk Metropolis-Hastings algorithm
  used in this study. $N_{stage}$ is limited to 2, and only the second
  half of the chain is used. As with 
  Algorithm \ref{alg:pt}, for pooled estimates,
  the likelihood at different Atwood numbers are combined by
  multiplying them together, whereas for unpooled estimates the
  procedure is repeated individually at each Atwood
  number.}\label{alg:pt}
~
\end{algorithm}

\renewcommand{\fgprfx}{Figs/ORIG-GIID-NORMAL}
\begin{figure}[h!]
\bc
\includegraphics[width=0.9\textwidth]{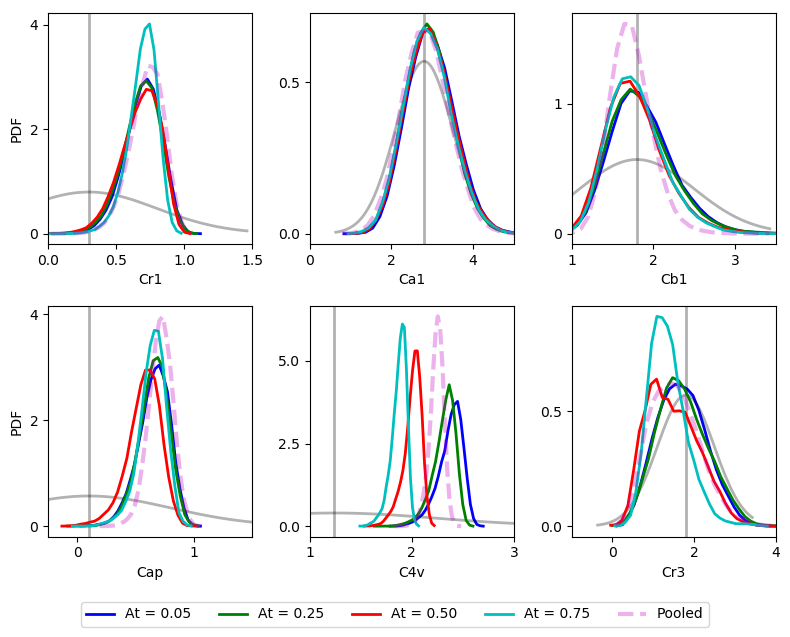}
\ec
\caption{Posterior distribution of the parameter values for the RANS
  model inferred in a Bayesian framework when calibrating the RANS
  model against DNS runs in pooled and unpooled scenarios. The colored
  curves are labeled in the legend. The prior distributions are shown
  in grey curves. Vertical grey lines are drawn at the values from
  \cite{schwarzkopf2016two}. Only marginal distributions are shown for
  convenience.}
\label{fig:unplldposterior}
\end{figure}
  
\begin{figure}[h!]
\begin{subfigure}{0.45\textwidth}
  \renewcommand{\fgprfx}{Figs/ORIG-GIID-NORMAL/Pooled}
  \includegraphics[width=\textwidth]{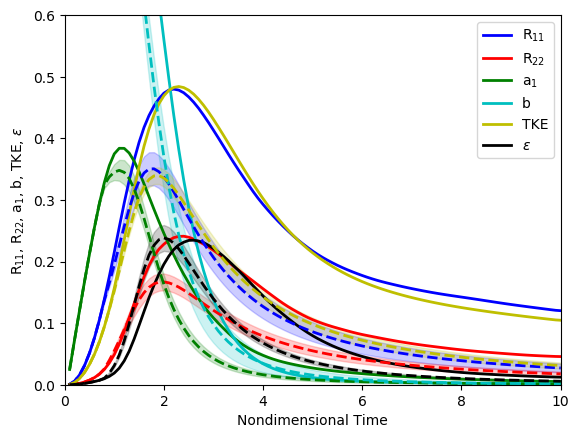}
  \caption{Pooled analysis}
\end{subfigure}
\begin{subfigure}{0.45\textwidth}
  \renewcommand{\fgprfx}{Figs/ORIG-GIID-NORMAL/1024_075}
  \includegraphics[width=\textwidth]{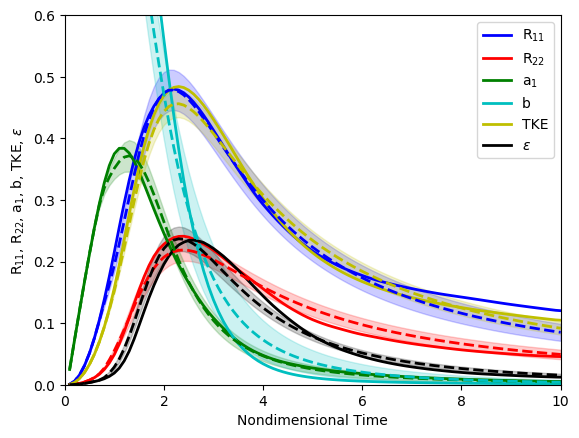}
  \caption{Unpooled analysis}
\end{subfigure}
\caption{Comparison of QoIs (between DNS and RANS at one Atwood
  (0.75) between pooled and unpooled analyses. As in Fig.~2, DNS
  results are shown in solid lines. Here the
  posterior distributions of parameter values are propagated through
  the RANS model to obtain the posterior distributions of the QoIs. The dashed
    lines now represent the mean over the posterior distribution of
    RANS estimates of the QoIs and the shading
    corresponds to $\pm$ one standard deviation. As
  expected the QoIs in the RANS simulations are seen to fit DNS data
  better in the unpooled analysis as compared to the fits in the
  pooled analysis.}
\label{fig:qoi}
\end{figure}

\subsection{Results from Pooled Bayesian Inference}
As discussed in Sec.~2, a single-point RANS turbulence model models
all turbulent phenomena in a local fashion. That is, each of the
turbulent processes is parameterized in terms of the local values of
relevant variables and their gradients. Thus, given such locality
assumptions, it is possible to imagine that the effects of Atwood
number would enter implicitly through the gradients of density and
gradients of other variables that are dependent on density. It may,
therefore, be anticipated that the coefficients themselves should not
depend on the Atwood number. Consequently, we first perform such a
calibration.

The marginals of the posterior distribution of parameters resulting
from such a calibration is shown in the heavy dashed magenta line in
the panels of Fig.~\ref{fig:unplldposterior}. It is seen in this
figure that the DNS data have moved the prior distributions---centered
at the point-estimate of \citet{schwarzkopf2016two}---significantly
away from that point estimate and greatly narrowed the distributions
from their prior values for three of the six parameters. In order to
ensure robustness of these inferences, the pooled analysis was
repeated with flat (uniform distributions with large bounds) priors;
the results did not change significantly and the results are not shown
to avoid redundancy.

The comparison between RANS and DNS estimates of the QoIs are shown
for only one of the Atwood numbers (0.75) in the left panel of
Fig.~\ref{fig:qoi}, again in order to limit the number of figures; large
differences are seen and this is true at other Atwood numbers as
well. The reader is referred to the lower five rows of Table~1 for the
actual size of the residuals at different Atwood numbers for the
different Bayesian calibrations considered. 

As discussed earlier, the large differences seen in the QoIs
leads to two possiblities: either that the model has serious
deficiencies or that the model has possibly a less serious deficiency
in being unable to simultaneously properly represent flows at
different Atwood numbers.

\subsection{Results from Unpooled Bayesian Inference}
In order to find out the origin of the model deficiency and its
significance, following the discussion in
Sec.~\ref{sec:pooled-unpooled}, we next consider unpooled or
independent calibration of the RANS model at each of the Atwood
numbers. The posterior distribution of parameters resulting from this
unpooled calibration is shown in solid lines in
Fig.~\ref{fig:unplldposterior}. As in the pooled analysis and
consistent with that inference, a significant shift away from the
prior distribution, both in terms of mean and variance, is seen in
these cases as well. Again, computations with a uniform prior
distribution with large bounds did not produce any significant
difference ensuring robustness of the inferences.

Comparisons of the RANS model fits to the DNS QoIs are shown in the
right panel of fig.~\ref{fig:qoi} (for the same Atwood number as
in the left panel). The fits are seen to be much better than
with pooled inference (left panel).

\begin{table}
\centering
\begin{tabular}{lcccccc}  
\toprule
Parameter & $C_{r1}$ & $C_{a1}$ & $C_{b1}$ & $C_{ap}$ & $C_{4v}$ & $C_{r3}$ \\
\midrule
JS-Divergence & 0.051 & 0.035 & 0.037 & 0.085 & 1.125 & 0.100 \\
\bottomrule
\end{tabular}
\caption{Janson-Shannon divergence for marginal distributions of RANS
  model at different Atwood numbers. $C_{4v}$, as compared to other
  parameters, have a strong distributional shift across Atwood
  numbers, suggesting a particular shortcoming of the RANS
  model: modeling the evolution of the turbulent length scale
  using (\ref{eq:ge5})}.
\label{tab:jsdiv}
\end{table}

\begin{figure}[h!]  \centering
\renewcommand{\fgprfx}{Figs/ORIG-GIID-NORMAL}
\includegraphics[width=0.6\textwidth]{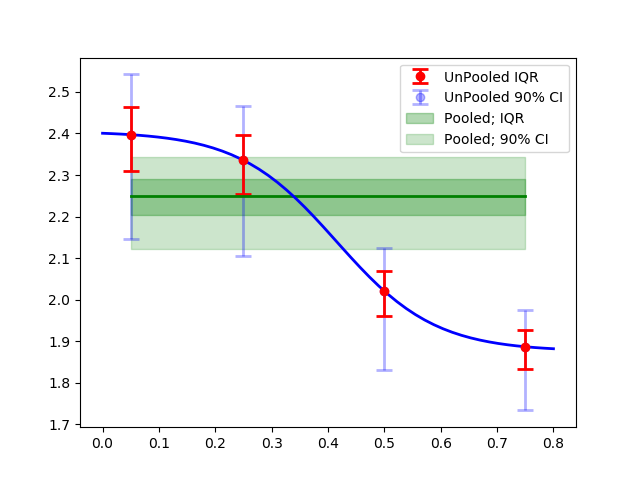}
\caption{Variation with Atwood number of RANS parameter C$_{4v}$ as
  inferred by Bayesian calibration}
\label{fig:c4vAt}
\end{figure}

\begin{figure}[h!]  \centering
\renewcommand{\fgprfx}{Figs/BSQRT-c66-NORMAL}
\includegraphics[width=0.9\textwidth]{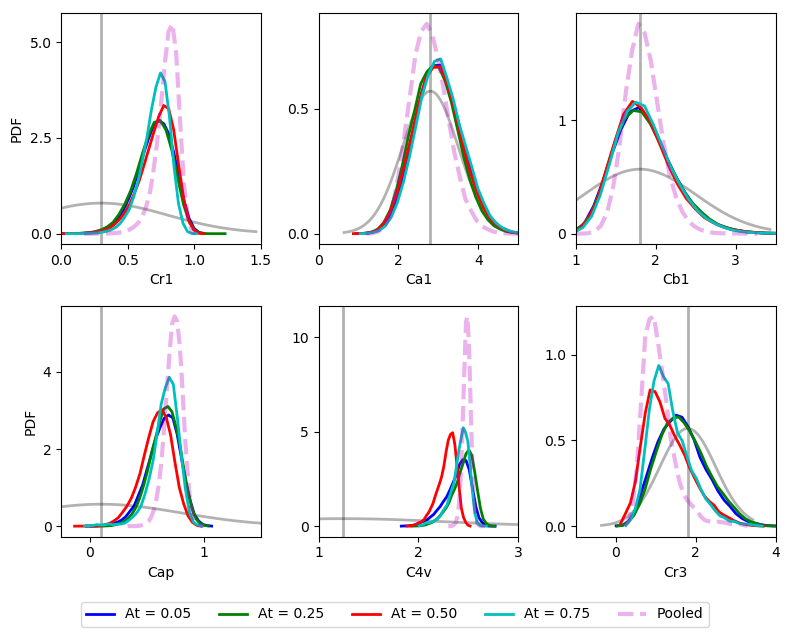}
\caption{Results from pooled and unpooled Bayesian analysis of the
  modified RANS model. For the first time, the modified model permits
  fitting of all Atwood number cases with a single set of parameter
  values (pooled). See Table 1. Results from earlier Bayesian analyses
  of the original RANS model is used to obtain the modification to the
  RANS model.}
\label{fg:modified-model-post}
\end{figure}

\begin{figure}[h!]  \centering
\renewcommand{\fgprfx}{Figs/SURR-GIID-UNIFORM}
\begin{subfigure}[t]{0.45\textwidth}
\includegraphics[width=\textwidth]{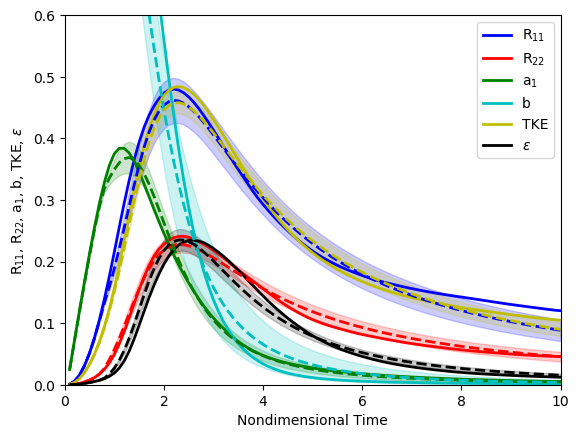}
\caption{Temporal trajectory of QoIs.}
\end{subfigure}%
\begin{subfigure}[t]{0.5\textwidth}
\includegraphics[width=\textwidth]{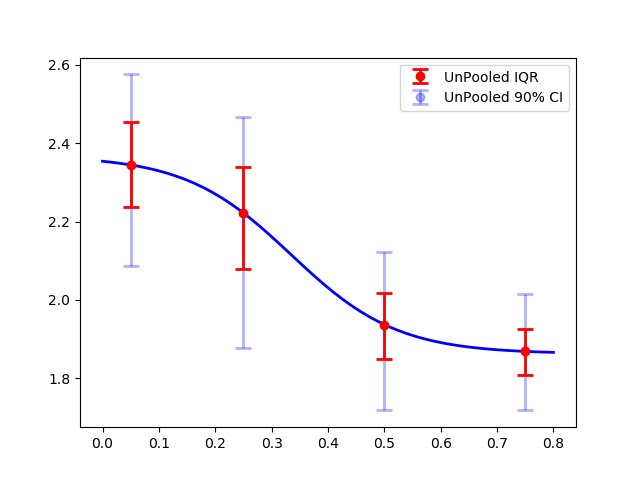}
\caption{Marginal distribution of $C_{4v}$ parameter vs. Atwood number}
\end{subfigure}
  \caption{Calibration with the RANS surrogate. The plots above are
    acquired by MCMC sampling on the response surface from the
    surrogate by the neural network.The temporal trajectories of QoIs
    closely resemble those acquired from the RANS model, and similar
    variations of the marginal distribution of the $C_{4v}$ parameter
    can be observed from sampling the surrogate.}
\label{fig:surr}
\end{figure}

\subsection{Variation of the Posterior Distribution of Parameters with
  Atwood Number and its Quantification}
Both the unpooled Bayesian inference and the unpooled point estimate presented earlier
 succeed in fitting the DNS estimates of the QoIs well. However, a significant
difference between the unpooled Bayesian and unpooled point-estimate
is that, while with the unpooled point estimate four of the six
parameters displayed large variations with Atwood number, it is seen
that with the unpooled Bayesian inference only one of the parameters
displays significant variations with Atwood number (lower middle panel
of Fig.~\ref{fig:unplldposterior} vs. red, cyan, and turquoise lines
in Fig.~\ref{fig:pointestimate-qois}).

While it is visually clear that distributions of $C_{4v}$ under
different Atwood numbers are distinctly different, we quantify the
distributional shift of the posterior PDF of model parameters
resulting from the change in Atwood number using the general
Jensen-Shannon divergence. The general Jenson-Shannon divergence is a
distance metric that measures the similarity between $n$ probability
distributions as:
\begin{align}
\text{JSD}_{\pi_1,\cdots,\pi_n}(P_1, \cdots, P_n) = H\Big(\sum_{i=1}^n\pi_i P_i\Big)P_i - \sum_{i=1}^n\pi_i H(P_i),\label{eq:jsdiv}
\end{align}
where $\pi_1,\cdots\pi_n$ are the weights for each probability
distribution $P_1,\cdots,P_n$. Here, $n=4$ for the four different Atwood numbers,
and we choose uniform weighting: $\pi_1=\pi_2=\pi_3=\pi_4=1/4$. $H(P)$ is the Shannon entropy for
distribution $P$. The Shannon entropy is defined as:
\begin{align}
    H(P) = -\sum_{i=1}^n P_i \log P_i
\end{align} 
The values are tabulated in Table. \ref{tab:jsdiv}.

The dependence of parameter $C_{4v}$ on Atwood number seen in
the lower middle panel of Fig.~\ref{fig:unplldposterior} is better
visualized in Fig.~\ref{fig:c4vAt}. A similar plot for the computations with a uniform
form for the priors shows little difference.

Next, we note that, after performing the above Bayesian analysis, in
order to further investigate the reason for the different dependencies
we found with the least-squares point-estimate and the Bayesian
inference, we considered the maximum likelihood estimate (MLE) which
uses the same likelihood function as in the Bayesian analysis.
Those estimates are shown in Fig.~1 and were discussed previously.

Finally, on a related note, in the probabilistic estimation context,
we show the marginal posterior distributions for ease of
visualization. However, the joint distribution can have a more
complicated structure than in one that may be imagined by composing
the marginal distributions alone. This is the case when, as is
typical, the turbulence modeling parameters are not all
independent. This is consistent with, and is borne out by, the fact
that none of the point estimates (see Fig.~1) happen to capture the
mode of the joint distribution that emerges from the Bayesian
analysis. Indeed, after the Bayesian analyses were conducted, a
limited amount of experimentation with ``basin-hopping'' extensions to
gradient-based point estimation methods did not help such an
``hybrid'' approach to capture the mode of the Bayesian analysis
either.  Such hybrid approaches, however, need to be further
investigated.

\subsection{Improving the Turbulence Model Based on Results of
  Bayesian Analysis}
As discussed above, pooled and unpooled Bayesian analyses of the RANS
model given a few DNSs at different Atwood numbers helped reveal a
dependence of Atwood number. That is, unlike with the point estimates,
large and systematic variations were confined to just one of the
parameters when Bayesian calibration was used.  While is is clearly
beyond the scope of this article to dwell on various optimization
stragegies and their respective advantages and disadvantages, we note
that in contrast to point estimates that typically rely on a local
search algorithm, a global sampling strategy that underlies the
Bayesian methodology is one of the reasons why the latter framework is
more robust. The flip side of this is a large added computational cost
and which we will address shortly. We now turn our attention
to how the results of the Bayesian analyses can be leveraged to
improve the turbulence model.

A straightforward way to do this would be to include or build in the
discovered variation of $C_{4v}$ with Atwood number. However, as
previously discussed, Atwood number itself is either a feature of the
initial condition or, at later times, a two-point or global
feature of the flow. As such, encoding the Atwood number dependence of
a parameter would be inappropriate in the one-point turbulence closure
model that is being considered. Therefore, we look into being able to
modify the turbulence model based on a local variable/correlation instead.

We discussed earlier that in this buoyancy-driven flow, the Atwood
number is a measure of the strength of the forcing. Next, towards
identifying local-variable proxies for Atwood number, consider the
temporal variation of various quantities in Figs.~2 and 4. It is
sufficient to consider the DNSs for this purpose (and confine
attention to any one panel in each of the two figures). In these
figures, it is seen (a) that the only non-zero variable/correlation in
the initial condition is $b$, (b) that the magnitude of $b$ increases
with At (Fig.~2 is at At = 0.50 and Fig.~4 is at At=0.75). These
observations suggest that $b$ may be a useful local proxy for Atwood
number. Consequently, we examine the variation of the maximum of $b$
(equivalently its initial value since it decays monotonically with
time) with Atwood number and find a quadratic dependence
($b_{max}\propto At^2$).

The identification of a local proxy for Atwood number paves the way
for making the transition to a structural modification of the
turbulence model starting from the discovered parameteric dependence:
we now consider a new term in (\ref{eq:ge5}) of the form 
\begin{equation}
\frac{2}{3}\frac{S_D}{\bar{\rho}K}\sqrt{b}a_1\bar{P}_{,1}.
\end{equation}

We repeat the pooled and unpooled Bayesian analysis of the modified
turbulence model. The posterior probability distributions of the
parameters are shown in Fig.~\ref{fg:modified-model-post}. First, the
systematic dependence of parameter $C_{4v}$ (lower-middle panel) on
Atwood number that is seen in the corresponding panel of
Fig.~\ref{fig:unplldposterior} is now absent. Second, no new
dependence is seen to be introduced in the unpooled
analysis. Furthermore, the RANS fits for the QoIs at each Atwood
number are good for both pooled and unpooled calibrations. This can be
seen in rows 8 and 9 of Table 1. While the good fits of the QoIs for
the unpooled analysis verify that that aspect of the original model
is retained, the modified turbulence model now allows, for the first
time, the pooled calibration to fit the QoIs at each of the four
Atwood numbers well (using a single set of parameters). This is a
significant improvement of the turbulence model.

\subsection{Surrogate RANS model using neural networks}
As mentioned earlier, a disadvantage of the Bayesian approach is the
added computational cost. The particular nature of the turbulent flow
that we chose to analyze in this article was such that the
RANS model could be integrated cheaply. As such, the added
computational cost involved in the Bayesian analysis was not much of an
issue. However, other turbulent flow settings may not be as forgiving
in that a single integration/realization of the RANS model could be
computationally expensive enough that the added computational cost of
the Bayesian analysis may be prohibitive. New computationally-efficient
sampling techniques may help alleviate this problem. However, we
consider a possible alternate stragegy---that of a surrogate RANS model.

Artificial Neural Networks (ANN) have been studied extensively for
their predictive capabilities in the context of machine learning, and
a feed-forward, multi-layered neural network has been shown to be
capable of approximating a continuous function arbitrarily well
\cite{cybenko1989approximation}. Given our current problem, the use of
an ANN for response surface exploration is of particular interest to
us; this subject has been well-explored \citep[e.g.,
see][]{anjum1997response}.

In this study, we use an ANN to predict the temporal trajectories of
turbulent correlation (see the RANS governing equations
(\ref{eq:ge1}-\ref{eq:ge5})) given the parameters $\bm\theta$. That is, we use
the ANN to establish a mapping from the RANS parameters to the
trajectories of tubulent correlations. Below, we briefly discuss some
of the main aspects of developing the surrogate model including data
compression, architecture of the neural network, and validation of
training quality using a held-out test set.

\paragraph*{Compression of the output vector:} While we are interested
in predicting the full temporal trajectories of the QoIs (e.g., the
trajectories shown in Fig. \ref{fig:qoi}), the temporal
trajectories contain a large number of time steps resultingly in a
high-dimensional output vector. At the same time, the smoothness of
these trajectories suggests temporal over-sampling and leads us to
consider principle components as a means of compression. Therefore, we reduce the
machine learning problem to predicting the PCA coefficients, and
recover/reconstruct the full trajectories using the PCA bases used for
the decomposition.  Since four PCA components are found to explain
in excess of 99\% of the variance, we restrict ourselves to learning four
components for each of five QoIs.

\paragraph*{Architecture of the Neural Network:} We use a deep fully-connected neural
network (DNN, a.k.a., multi-layer perceptron) with 10 hidden layers and with
each layer consisting of 30 neuron units. The size of the input layer
corresponds to that of the  parameter vector and given the compression
discussed above, the size of the ouput layer is 20.

\paragraph*{Validation on test set:} We evaluate the performance of our
neural network by using a held-out test set. The size of the test set
is held fixed at 10\% the size of the training set and both sets are
sampled from the same distribution. 

\paragraph*{Results:} After training the ANN, we conducted unpooled
Bayesian calibration of the RANS model, but now using the the DNN-RANS
surrogate instead of integrations of the actual RANS
model. Representative sample results are shown in
Fig.~\ref{fig:surr}. With the DNN-surrogate, it is seen that not only
are the fits to the DNS estimates of the QoIs good, but that the
variation of the $C_{4v}$ parameter with Atwood number is also well
approximated. 

In order to avoid a long digression, we skip details, but note that
convergence of a Markov chain to the target posterior distribution is
typically slow. For this reason, chains of length a million or more
samples are typically used to ensure robustness of inferences. For the
training/testing phase of the DNN-surrogate, we used a set of 500
integrations of the RANS model. After the DNN-surrogate has been
trained it is fairly cheap to evaluate it. Thus in a situation where
individual RANS integrations are costly this strategy can lead to
significant computational savings. We do not dwell on actual
differences in computational costs in this proof-of-principle
demonstration for the reason that the RANS model itself is presently
computationally cheap. We expect the very low end of cost
advantages of the surrogate approach in cases where individual RANS
integrations are expensive to be ten or more; more
realistic estimates require actually working through such a case.

\section{Discussion and Conclusion}\label{sec:disc}
In this study, we considered various methods of analyzing and
calibrating an approximate model of a complex multiscale sytem given a
few well-resolved simulations. We first
considered various point-estimate approaches. These ranged from using
previously published point estimates, to new ones that used ordinary
least squares and maximum likelihood estimates. While it was found
that some of the unpooled point estimates were able to fit the QoIs well,
all pooled point estimates and some unpooled point estimates produced
poor estimates of the QoIs. We rationalized the behavior of the point
estimates by appealing to the effects of pooling versus unpooling of
the different Atwood number cases and the effects of considering or
not considering the dynamical nature of the approximate model.

Next we considered the behavior of the parameters themselves in the
cases that produced good estimates of the QoIs. Not only were the
parameter estimates themselves different, depending on the method
used, but their variation with Atwood number was also very
different. For these reasons, the main insight into the approximate model
provided by the various point estimates was limited to indicating that
the model under consideration contained enough of the relevant
phenomenology to properly represent the particular flow that we
consider, but that they had to be individually calibrated.  While
valuable, it did not point in further specific ways how the model
could be improved.

We next conducted Bayesian analysis of the approximate model using
the same data as was used for the point estimates. Not surprisingly,
pooled and unpooled Bayesian analyses first verified and confirmed the
insight provided by the point estimate approaches. That is that while
the approximate model could produce QoIs that compare well with DNS
estimates at each Atwood number, they are unable to do so using a
single set of, now probabilistic or interval estimates. Next, however,
this approach provided further specific pointers towards how this
shortcoming could be overcome. Compared to the point estimates where
different variations with Atwood number were produced with different
variants of the estimators, the Bayesian analyses produced similar
posterior distributions at all Atwood numbers for all but one of the
model parameters.

The minimization of parameter variability with Atwood number may be
thought of as the key contribution of the Bayesian analyses as far as
insights into the workings of the approximate model is
concerned. This is because, in contrast to the wide range of parameter
variations seen in point estimates, the systematic variation of the
posterior distribution of a minimal set of parameters (in this case, a
single one) suggests that the former is likely an artifact of the
point-estimate approach to calibration. Furthermore, a small set of
systematic parameter variations allows for an immediate improvement of
the model (as we demonstrate), with initial improvements being based
purely on statistical grounds. However, such improvements could then
foster better physics- or dynamics-based improvements. Given the
non-uniqueness of closures, if multiple such improvements should
arise, then it is possible that one can use the principle of Occam's
razor in conjunction with plausibility to choose from among them
\cite{farrell2015bayesian}.  

Next, we consider two aspects of computational cost associated with
the procedure that we have considered and find useful: first the
relative difference in cost between point estimates and probablistic
estimates and next, the cost of the approximate model itself. Not surprisingly,
probablistic estimates when performed using sampling are costlier than
point estimates. However, this difference in cost can be greatly
reduced by using variational methods of probabilistic inference, an area
that has seen major improvements in the last couple of years. We
present a few details below.

When using random-walk (Metropolis-Hastings) based algorithms for
generating members of the Markov chain, the stepsize of MCMC scales
inversely as the dimension of $\vtheta$ (number of parameters being
inferred): if h is the stepsize, $h \sim 2.4^2/$dimension(\vtheta)
\cite{gelman1996efficient}. Consequently, when the dimension of $\vtheta$ is
large, a larger number of MCMC steps have to be taken so that cost
scales as $dim(\vtheta) \times C_{step}$ where $C_{step}$ is the
computational cost of a step. Thus, for a reasonable chain length, the
computational cost of the MCMC-based scheme scales as, say, 
$10^3\times dim(\vtheta) \times C_{step}$.  However, by similar
straightforward scaling arguments, the cost of a point estimate scales
only as, say, $10^2 \times C_{step}$, leading to the probabilistic estimate
being costlier by a factor of about $10 \times dim(\vtheta)$.  So for
$dim(\vtheta) \sim 10$, as in the present study, the cost difference
is about 100 and the study was computationally feasible. However, this
becomes a problem when $dim(\vtheta)$ is large. 

Recent improvements in computational inference methods, however, can
be brought to bear on this issue. For example, computational cost of
Hamiltonian Monte Carlo only scales as, say, $10^3 \times C_{step}$
\cite[e.g., see][]{beskos2013optimal}. Even better, variational
inference methods have now been developed that bring the cost of
probabilistic inference to levels comparable to that of point
estimates \cite[e.g., see][]{blei2017variational}. However, it has to
be noted that with variational inference, an additional error
is introduced since optimal parameters for the source
distributions that best match the actual posterior distribution will
be found, rather than the actual posterior distribution itself as we
currently compute. Nevertheless, we think that the ideas presented in this
article can be scaled up to larger problems by the use of
probabilistic inference (but with caveats
such as the one mentioned above).
 
Next, we consider the issue of computational cost of one step in the
inference procedure, viz. $C_{step}$ above. It is dominated by the
cost of evaluating the likelihood function which in turn requires an
integration of the approximate model. If an individual integration of the approximate model itself
is computationally intensive (while still being orders of magnitude
cheaper than well-resolved simulations) then conducting either point
estimate based analysis or probablistic estimate based analysis can be
computationally expensive.  Towards reducing such costs, we explored
the use of an artificial neural network as a surrogate model for the
RANS solver and we demonstrated the feasibility of using such a
ANN-based surrogate in conducting the Bayesian analysis: the Bayesian
analysis using the ANN-surrogate recovered similar parametric
dependencies as the Bayesian analysis that was performed using RANS
integrations. Thus in a situation where individual integrations of the
approximate model are themselves costly we anticipate that this
strategy can lead to significant enough computational savings that the
Bayesian analysis can be performed.

In other considerations, both the well-resolved model and the approximate model
were deterministic in the case we chose to illustrate the process of
improving the approximate model using probabilistic or interval inference. However,
since the methods that we use are agnostic about the deterministic or
stochastic nature of either the well-resolved model or the approximate model,
application of this methodology to situations wherein either the
well-resolved model or the approximate model is stochastic is easily achieved on taking
into consideration prior information available about the nature of the
stochasticity involved. That is, for example, we anticipate that the
procedure for  model improvement should work in the setting of the Markov
State Model or other such models in the context of data coming from either a
deterministic or stochastic, but well-resolved model. We also note
that we have leveraged certain aspects of Bayesian estimation to serve
the pupose of model improvement. It remains to be seen if other
aspects resulting from Bayesian estimation, such as ensemble
characteristics, can also be exploited for similar purposes.

Bayesian analysis can be used in many ways to help with the task of
reduced-order modeling but, the fundamental aspect of Bayesian
analysis that permits its use in such a fashion is related to its
ability to create knowledge \cite[e.g.,
see][]{babuska2004verification, edeling2014bayesian,
  edeling2014predictive, farrell2015bayesian}.  A few recent studies
have highlighted such uses and include using Bayesian approaches to
estimating and characterizing errors and uncertainty in RANS models
and comparing and selecting from among different models. We add to
this nascent body of literature by showing how Bayesian analysis can
be leveraged to improve models themselves: the pooled and unpooled
Bayesian analysis we conducted revealed unanticipated parameteric
dependencies, and then we closed the analysis-improvement loop by
using the discovered dependency to effect a structural modification of
the model that removed the dependency while not introducing others and
in a fashion that is consistent with the modeling approach \cite[see
also][]{nadiga2019improved, degennaro2018model}. Finally, we
anticipate that this methodology will also be useful in improving
reduced-order models when such models include other
parameter-dependent model terms that serve to either stabilize or
improve the fidelity of the model.

\section*{Acknowledgements} We thank the referees for their extensive
comments and suggestions. The presentation of this article has
benefitted greatly from them. This research was funded by the US
Department of Energy, in part under the Laboratory Directed Research
and Development program at the Los Alamos National Laboratory (LANL), and in
part by the Mix and Burn research initiative under the Physics and
Engineering Models (PEM) component of the Advanced Simulation and
Computing Program (ASC) at LANL.

\section*{References}
\bibliographystyle{plainnat}
\bibliography{nl16}

\end{document}